# Lognormal distribution of firing time and rate from a single neuron?

Eszter A. Kish [1], Claes-Göran Granqvist [2], András Dér [3], Laszlo B. Kish [4,a]

[1] *Center for Learning and Memory, The University of Texas at Austin, 1 University Station, Stop C7000, Austin, TX 78712-0805*

[2] *Department of Engineering Sciences, The Ångström Laboratory, Uppsala University, P. O. Box 534, SE-75121 Uppsala, Sweden*

[3] *Institute of Biophysics, Biological Research Centre of the Hungarian Academy of Sciences, Temesvári krt. 62, P.O.B. 521, Szeged, H-6701, Hungary*

[4] *Department of Electrical and Computer Engineering, Texas A&M University, College Station, TX 77843-3128, USA*

**Abstract**. Even a single neuron may be able to produce significant lognormal features in its firing statistics due to noise in the charging ion current. A mathematical scheme introduced in advanced nanotechnology is relevant for the analysis of this mechanism in the simplest case, the integrate-and-fire model with white noise in the charging ion current.

In a recent review by Buzsaki and Mizuseki [1] the wide occurrence of lognormal-like distributions in the structural organization parameters and the firing rate of neurons were surveyed and their assumed functionalities were explored. It was assumed that the lognormal distribution of firing rates is the consequence of the specially organized circuit connectivity and the high complexity of the nervous system. Subsequently a preprint with additional study has been published by Scheler [2].

The natural question emerges if the internal dynamics of single neurons is already able to produce a lognormal firing feature due to its inherent stochastic features.

At the first look, such assumption looks rather unconventional. For example, several works study stochastic resonance with additive Gaussian noise [3,4] *in the membrane potential*. Due to the level-crossing properties of Gaussian noises, such models obviously result in a distribution of firing rates with no long-tail but exponential cutoff. Moreover, in models with bifurcation theory [5] and stochastic Hodgkin-Huxley channels [6,7], interesting implications of spontaneous stochasticity were found however the papers do not mention lognormal dynamics.

[a] Corresponding author. Laszlokish@email.tamu.edu ; phone: +1-979-847-9071

Still, experimental observations of lognormal firing statistics on lower levels of hierarchical organizations [8] seem to justify the question. Below, we present a quantitative example how the combination of plausible statistical assumptions and the simplest neuron model can lead to the appearance of lognormal firing rate distribution on the level of single neurons.

One of the well-known mathematical ways that lognormal distribution is obtained is a random walk on an axis with logarithmic scale (geometric random walk) resulting a growing Gaussian distribution over the axis, which is (due to the exponential stretch) equivalent to a lognormal distribution on the linear scale. Relevant applications of this model are stochastic stone cracking with fixed mean cracking fraction or its inverse process via coagulation/aggregation of nanoparticles [9]; both models result in lognormal size distribution.

However, these old models cannot account for the lognormal distribution of nanoparticle sizes at advanced vapor based fabrication methods [10,11] where the growth is condensational (linear in time) and when coagulation/aggregation is avoided. The origin of lognormal distribution in such cases was explained by a lognormal residence time distribution in the growth zone (vapor zone) of nanoparticle fabrication. Proceeding through the zone with a Brownian motion superimposed on a constant drift velocity results in a lognormal-like residence time distribution whenever the drift is strong and the starting point of the zone has a reflecting boundary [10,11]. The discrete difference equation describing the progression though the zone is given as:

$$x(k) = x(k-1) + \delta + \zeta(k)\sqrt{D} ,  \qquad (1)$$

where $k$ is discrete time (measured in computational steps); $x(k)$ is the position coordinate of the growing particle, $\delta$ is the drift velocity; $\zeta(k)$ is a random number with Gaussian (or other fast-cut-off, such as uniform) distribution, zero mean value, and unity variance; and $D$ is the diffusion coefficient, which is the mean-square of the velocity noise resulting in the random-walk component superimposed on the drift. When the $\zeta(k)$ random numbers are independent, $\zeta(k)$ represents a band-limited white noise thus the resulting random walk component is a Brownian motion.

The motion described by Equation 1 begins at $x(0) = x_0$ and the first-passage time to the other end $x_{th}$ of the zone is a random variable $k_{th}$. When the $x_{th} \leq x(k_{th})$ is first satisfied, the growth process stops and $k_{th}$ is recorded thus $k_{th}$ is the residence time in the growth zone, that is, time spent by the linear growth. Here the threshold coordinate is given as $x_{th} = x_0 + L$, where $L$ is the length of the growth zone. The starting point $x_0$ is a reflecting boundary, that is, the $x_0 \leq x(k)$ condition is enforced during the whole motion. The condition of strong drift means that the drift is greater than the critical value $\delta_0$:

$$1 < \delta / \delta_0 ,  \qquad (2)$$

where the critical drift depends on the strength of the noise and the length of the zone:

$$\delta_0 = \frac{D}{L}. \tag{3}$$

In the case of $\delta = \delta_0$, the noise-free drifting time through the system is equal to mean first passage time due to the noise at zero drift. At strong drifts (Equation 2) the set $\{k_{th}\}$ of residence time distribution is lognormal and, because the particle size is a linear function of the residence time, lognormal particle size distribution is the result, see Figure 1.

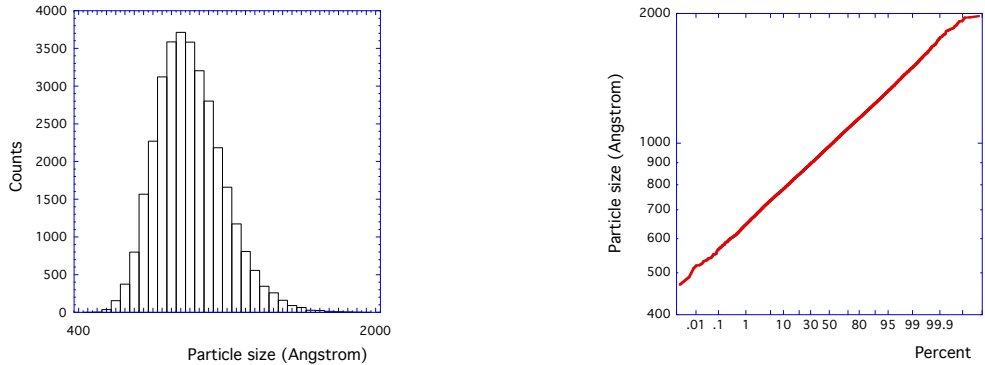

**Figure 1**. Histogram of density function (left), and cumulative distribution in log-Gaussian plot (right) of the sizes of 100 thousand nanoparticles by condensational growth, without coagulation, due to Brownian motion superimposed on linear drift in the growth zone (based on [10,11]). The log-Gaussian plot is much more efficient than the histogram to follow the behavior in the long tail and a straight line represents ideal lognormal distribution. Drift: 16.6 times the critical drift.

To explain the observed lognormality in the single protein molecule detection scheme with fluorescent quantum dots, the same mathematical model was applied for quantum-dot-marked-molecules drifting in a nanofluidic channel through a zone exposed to a laser beam. Even the additional photonic shot noise could not destroy the lognormal feature in the size distribution of photon bursts [12], see Figure 2.

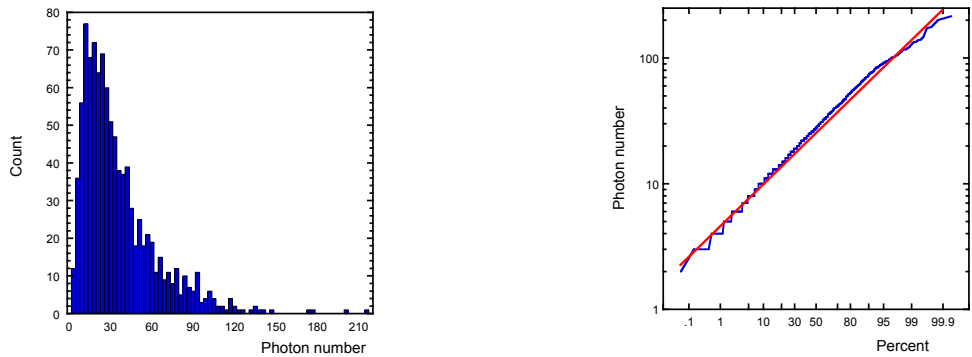

**Figure 2**. Histogram of density function (left), and cumulative distribution in log-Gaussian plot (right) of photon burst sizes in single molecule detection with quantum dots [12]. Even the additional photon shot

noise in the model is unable to destroy the lognormal characteristic. Drift: 1.9 times the critical drift.

There is a striking similarity between the model described above and the integrate-and-fire model, the simplest dynamical neuron model, if we suppose that there is a band-limited white noise in the ion current, see Figure 3 for its circuit representation.

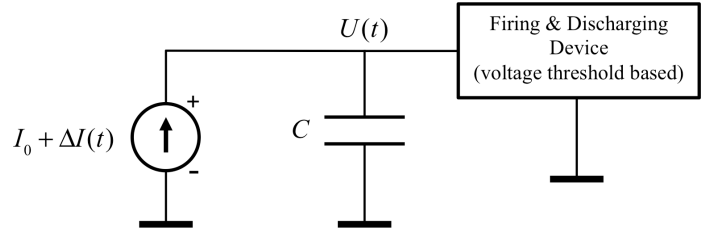

**Figure 3**. Circuit representation of the integrate-and-fire model: a capacitor is charged by a current generator from the initial potential level $U_0$ up to the threshold potential $U_{th}$ where the firing takes place and the capacitor is discharged. In the noise-free case, the membrane potential $U(t)$ is drifting with $\delta = I_0 / C$ velocity up to the firing threshold, where $I_0$ is the charging ion current and $C$ is the capacitance. The current noise $\Delta I(t)$, when it is a band-limited white noise with Gaussian or other amplitude density of fast cut-off, results in the sum of Brownian motion and a linearly drift in the membrane potential $U(t)$. With a reflecting boundary at the initial potential value (or proper amplitude density of the noise to prohibit backward propagation events) this is the same mathematical model as the one leading to Figure 1 (see Equations 1-3).

Thus it is straightforward to apply the model as follows. In the discrete-time model, the coordinate of the motion is the membrane potential $U$, the drift velocity of potential is $\delta$, and $D$ is the mean-square of the noise in the ion current:

$$U(k) = U(k-1) + \delta + \zeta(k)\sqrt{D} ,\qquad(4)$$

where k and $\zeta(k)$ are defined in the same way as in Equation 1. In accordance with Equations 2 and 3, the critical drift is given as:

$$\delta_0 = \frac{D}{U_{th} - U_0} ,\qquad(5)$$

where the initial potential value is $U_0 = U(0)$ and the potential threshold of firing is $U_{th}$. The starting point $U_0$ is a reflecting boundary, that is, the $x_0 \leq x(k)$ condition is enforced during the whole process. When the $U_{th} \leq U(k_{th})$ is first satisfied, the neuron fires, the membrane potential is discharged and the whole charging process starts from the beginning. The actual $k_{th}$ value is recorded; it is the time interval between the former and the present firing events (inter-spike interval). Here we assumed that the firing/discharging process is negligibly short compared to the inter-spike interval. because Equations 1 and 4 and the mathematical conditions are identical, in the strong drift limit (see Equation 2), the set $\{k_{th}\}$ has obviously lognormal distribution. Furthermore, because any power function of a lognormally distributed random variable is also lognormal, not only the inter-spike intervals but also the firing frequency will have lognormal-like distribution if the firing/discharging process is negligibly short compared to the inter-spike interval.

Figure 4 shows the histogram obtained by computer simulations of the integrate-and-fire model with Equation 4 with $U_0 = -60$ mV, $U_{th} = -40$ mV, and relative drifts $\delta/\delta_0 = 6$ and $24$, respectively. Both the time and frequency data show the familiar skewed shape.

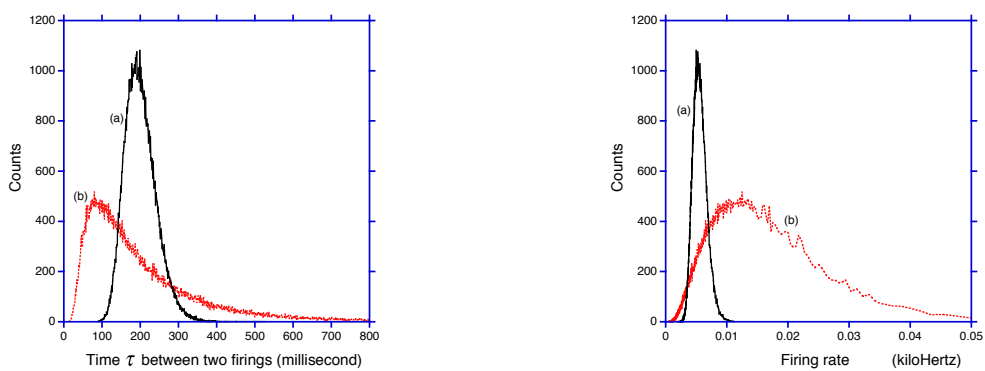

**Figure 4**. Computer simulations of the integrate-and-fire model with white noise in the ion current causing a random walk (Brownian motion) superimposed on the linear drift of the potential. The same random walk model with special parameters used as in getting Fig. 1. The width and skewness of the resulting lognormal-like distribution depend on the relative drift, which is the drift normalized to the critical drift value. Because any power function of a lognormally distributed random variable has also lognormal distribution, the lognormal distribution of time intervals between firing implies a lognormal distribution of firing frequency (in the limit when the time spent for firing/discharging can be neglected). Drift (a) 6 times and (b) 24 times the critical drift.

It is open question if the additive noise in the ion current is strong enough to yield the observed distribution of firing frequency of single neurons. However, models and observations [13] regarding the stochastic closing and opening of ion channels indicate that the noise can be sufficiently strong. It is also an open question and subject of future studies how much does the distribution deviate from lognormal in those cases when the noise spectrum is 1/f [14,15] instead of white and in the case of more advanced neuron models.

Finally, we note that Longtin [16] studied stochastic resonance phenomena in the time distribution of firing events at sinusoidal excitation of the Fitzhugh-Nagumo neuron model. To introduce stochasticity, a white noise term was added to the time derivative of the potential. In the case of no sinusoidal excitation, a skewed density function (resembling lognormal) of the time intervals between firing events can be seen. However, this fact was not commented because it was considered only as the base line of observations and the paper was focusing on the induced periodicity and stochastic resonance at sinusoidal driving in the presence of noise.

**Acknowledgements**

Discussions with Sergey Bezrukov (NIH) and György Buzsáki are appreciated.